# Firefly-Algorithm Supported Scheme to Detect COVID-19 Lesion in Lung CT Scan Images using Shannon Entropy and Markov-Random-Field

**Venkatesan Rajinikanth[1], Seifedine Kadry[2,\*], Krishnan Palani Thanaraj[3], and Krishnamurthy Kamalanand[4], Sanghyun Seo[5]**

[1] Department of Electronics and Instrumentation Engineering, St. Joseph's College of Engineering, Chennai 600119, India; v.rajinikanth@ieee.org

[2] Department of Mathematics and Computer Science, Faculty of Science, Beirut Arab University, Lebanon, skadry@gmail.com

[3] Department of Electronics and Instrumentation Engineering, St. Joseph's College of Engineering, Chennai 600119, India; palanithanaraj.k@gmail.com

[4] Department of Instrumentation Engineering, MIT Campus, Anna University, Chennai 600044, India; kkamalresearch@gmail.com

[5] School of Computer Art, College of Art and Technology, Chung-Ang University, Anseong-si 17546, Kyunggi-do, Korea; sanghyun@cau.ac.kr

\* Correspondence: skadry@gmail.com;

**Abstract:** The pneumonia caused by Coronavirus disease (COVID-19) is one of major global threat and a number of detection and treatment procedures are suggested by the researchers for COVID-19. The proposed work aims to suggest an automated image processing scheme to extract the COVID-19 lesion from the lung CT scan images (CTI) recorded from the patients. This scheme implements the following procedures; (i) Image pre-processing to enhance the COVID-19 lesions, (ii) Image post-processing to extract the lesions, and (iii) Execution of a relative analysis between the extracted lesion segment and the Ground-Truth-Image (GTI). This work implements Firefly Algorithm and Shannon Entropy (FA+SE) based multi-threshold to enhance the pneumonia lesion and implements Markov-Random-Field (MRF) segmentation to extract the lesions with better accuracy. The proposed scheme is tested and validated using a class of COVID-19 CTI obtained from the existing image datasets and the experimental outcome is appraised to authenticate the clinical significance of the proposed scheme. The proposed work helped to attain a mean accuracy of >92% during COVID-19 lesion segmentation and in future, it can be used to examine the real clinical lung CTI of COVID-19 patients.

**Keywords:** COVID-19 pneumonia; CT scan image; Firefly algorithm; Shannon entropy; Markov-Random-Field segmentation; Evaluation.

---

## 1. Introduction

The recent developments in science and technology helped the human community to get improved medical facility for disease detection and treatment execution process. Further, most of the life threatening diseases in humans are controlled due to prescreening and vaccination procedures. Even though a substantial amount of preventive methods are taken; the spread of the infectious diseases are rising globally due to various circumstances and the communicable illness called the Coronavirus Disease (COVID-19) is one such disease; causing severe impact globally [1-3]. Because of its severity; the infection rate due to COVID-19 is growing rapidly and has affected large human communities irrespective of their age, gender and race [4-6].

The cause of COVID-19 is due to a virus named Severe Acute Respiratory Syndrome-Corona Virus-2 (SARS-CoV-2) and this disease originally originated in Wuhan, China, in December 2019 [7]. Outbreak of this disease has emerged as a global challenge and a significant number of research works are initiated to discover the solution to control the infection rate and the spread of COVID-19 [8-10]. Further, the recent research works related to; (i) disease detection [11-16], (ii) treatment planning [17-21] and (iii) disease prediction [6,22,23] has helped to get insight regarding COVID-19.

The earlier research and the clinical detection revealed that, COVID-19 causes infection in the respiratory tract and initiates severe pneumonia. Further, the research also revealed that, the early symptoms of COVID-19 are subclinical and it requires a dedicated clinical procedure to detect and confirm the disease. The common clinical level diagnosis involves in; (i) Sample assisted analysis of COVID-19 using Reverse Transcription-Polymerase Chain Reaction (RT-PCR) and (ii) Image assisted evaluation using lung Computed-Tomography (CT) scan images and/or Chest radiographs (Chest X-ray). During the disease screening process, biological samples are initially collected from the patient and the RT-PCR test is performed. When the test result is positive, then the patient is admitted into the hospital for the further diagnosis and treatment. When the patient is admitted for treatment, the physician will initiate the treatment process to cure the patient using the prescribed treatment procedure which will reduce the impact of the pneumonia [24-26].

For the COVID-19 pneumonia patients, the pulmonologist will suggest a sequence of investigative tests to identify the origin, location, and severity of the pneumonia. Initially, the preliminary tests, such as blood count, blood-gas examination and pleural-fluid assessment [13] are performed. The image supported practices are also regularly performed to trace the infection in lung, which can be additionally inspected by a skilled doctor or a computerized arrangement to recognize the harshness of the pneumonia. Compared to the chest X-ray, the CT scan Image (CTI) is commonly used due to its merit and the three-dimensional view support. The recent research work published on COVID-19 also verified the advantage of the CT in detecting the infection in the respiratory tract and pneumonia [20-25].

Recently, a number of methodologies are implemented for the clinical level detection of COVID-19 using the RT-PCR and the imaging methods [11-13]. Most of these earlier works are combined the RT-PCR along with the imaging procedure to confirm and treat the disease [9,18-20]. The CTI is commonly used as the prescribed modality and the axial as well as the coronal views are analyzed to detect the severity of the infection and based on the outcome, effective treatment is suggested to treat the disease with suitable drug. The recent work of Rajinikanth et al. [27] developed a computer assisted methodology to assess the COVID-19 lesion using the lung CTI of coronal-view using a series of procedures. This work implemented few operator assisted steps to achieve superior outcome during the COVID-19 evaluation.

The intention of the proposed research is to develop a computerized scheme to extract and evaluate the COVID-19 lesions from the lung CTI. The recent studies on COVID-19 pneumonia also verify that, the estimation of infection severity level is necessary to arrange for a suitable treatment plan to cure the patients. In a real-time situation, lung infection due to COVID-19 initiated pneumonia is evaluated by a skilled doctor and based on the suggestion; probable healing actions are suggested and implemented to reduce the infection level. The personal recognition using a doctor will be a tedious job when the number of patients is more and hence, it is essential to propose a computerized image examination practice to assist the physician during the pneumonia diagnosis. This scheme will significantly lessen the diagnostic burden of the physician and will assist the physician to track the changes in the lesions (reduction/progress in infection) when the treatment is implemented.

The aim of this research is to develop a computerized COVID-19 lesion detection system to reduce the diagnostic burden of the pulmonologist responsible for the investigation of the CTI to discover the pneumonia level in lungs. The proposed research executes a sequence of techniques, such as artifact removal, Firefly Algorithm and Shannon-Entropy (FA+SE) based multi-thresholding, Markov-Random-Field (MRF) segmentation and validation of the proposed system using a comparison with respect to the Ground-Truth-Image (GTI). In this work, the COVID-19 dataset available in [28] is considered for the initial assessment. This dataset consists of one hundred numbers of COVID-19 images with dimension 512x512x1 pixels and all the images are associated with the GTI. In the proposed work, 50 images of the benchmark dataset are considered for the assessment and the average accuracy of >92% is attained with the proposed technique. The performance of the proposed system is finally validated using the clinical grade CTI of the Radiopaedia database [29] and achieved better result.

This research is arranged as follows; section 2 presents the methodology and section 3 outlines the attained results. Section 4 and 5 presents the discussions and the conclusion of the present research work.

2.  **Methodology**

This section of the research presents the methodology implemented to detect the lung lesions due to COVID-19. The structure of the proposed examination procedure is depicted in Figure 1.  Initially, the essential two-dimensional (2D) slices of the CTI (gray scale) are collected for the examination. The organ of interest is the lung and in the CTI, the lung is associated with various artifact sections, such as the heart, bone and other body parts and these artifacts are to be removed for the efficient assessment of lung and its infection. In this research, the stripping methodology discussed by Palani et al. [30] and Rajinikanth et al. [31] is executed to remove the artifacts with a bi-level threshold filter. The separated lung segment is then considered for image pre-processing and post-processing methodology. The image pre-processing step implements a three-level threshold to separate the image into; the background, normal lung section and the pneumonia lesion. In this work, the image enhancement is achieved using the Firefly-Algorithm (FA) guided Shannon-Entropy (SE) thresholding. The enhanced pneumonia lesion in the lung is then extracted by implementing Markov-Random-Field (MRF) segmentation. Finally, the extracted COVID-19 lesion is compared against the Ground-Truth-Image (GTI) and based on the attained performance values; the superiority of the proposed image processing tool is confirmed.

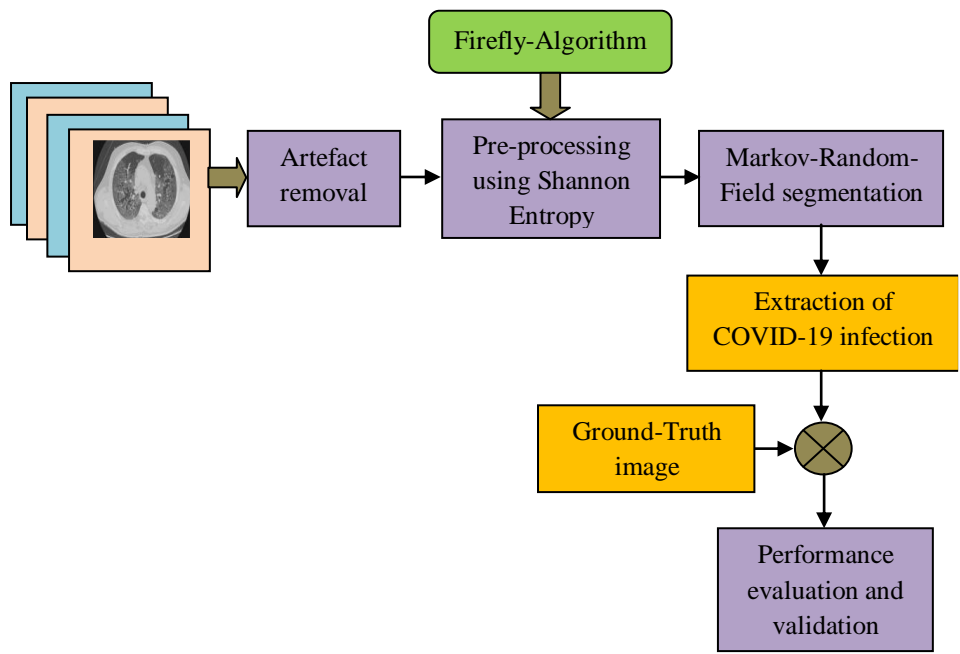

**Figure 1.** Different operations involved in the proposed COVID-19 lesion detection system

## 2.1 COVID-19 image database

The pneumonia infection due to COVID-19 is generally appraised in hospitals using the chest X-ray/CTI [9-12]. The visibility of CTI is better compared to chest X-ray and hence, this research considers the CTI for the experimental investigation. The COVID-19 is a novel disease and the availability of the images for the research is very limited due to various constraints. This work considered the COVID-19 pneumonia image dataset of [28] to test the performance of the proposed image processing method. This database consists of 100 CTI with dimensions 512x512x1 pixels and all the existing images are associated with related Ground-Truth-Images (GTI). This research considered 50 numbers of 2D slices recorded with the axial-view for the assessment and the sample images considered in this research is depicted in Figure 2.

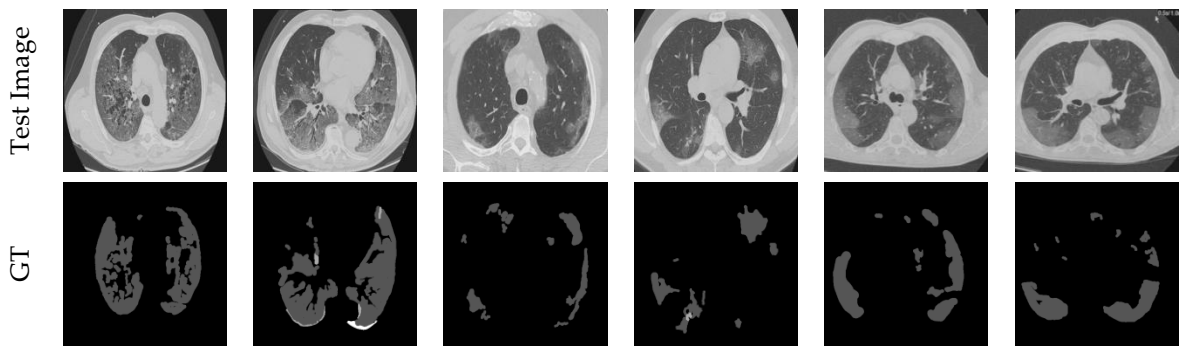

**Figure 2**. Sample test images of the benchmark COVID-19 pneumonia database

The initial level of investigation of COVID-19 lesions is performed using the considered 50 numbers of CTI. After confirming its performance based on a comparative assessment with the GTI, its significance is then validated using the clinical grade images attained from the Radiopaedia database and the sample trial images of this database is depicted in Figure 3. During the validation practice, 45 numbers of axial-view 2D CTI (3 patients x 15 slices = 45 slices) are considered and the attained results are evaluated.

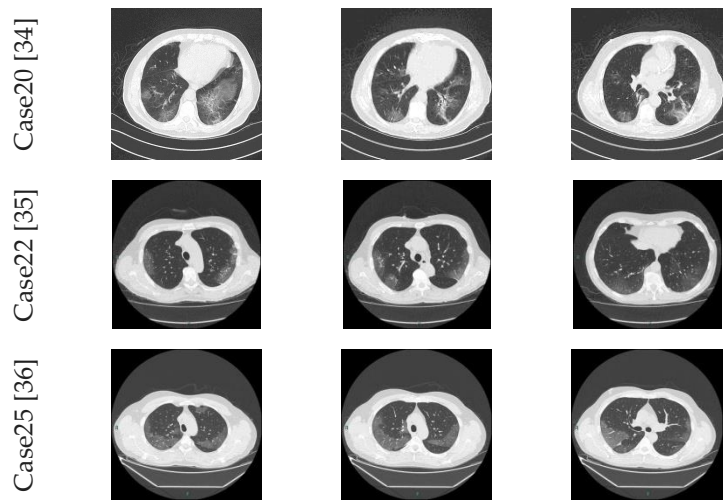

**Figure 3.** Sample test images attained from the Radiopaedia case-studies

### 2.2 Image pre-processing

The artifact elimination is the initial procedure implemented to separate the lung region from other sections using the artifact-elimination procedure discussed in [30, 31]. Later, the visibility of COVID-19 infection in the lung is enhanced by implementing the pre-processing procedure based on the Firefly-Algorithm (FA) assisted threshold technique discussed in [31]. The enhancement attained with the Tsallis-Entropy based tri-level thresholding is very poor and hence the Shannon's Entropy (SE) thresholding is implemented to improve the pneumonia infection due to COVID-19 [32]. The entropy assisted image processing is normally considered to identify the abnormalities in the image for further examination.

The SE considered in this work is discussed below;

Let, $AxB$ depicts a gray scale image of a chosen dimension and $f(X,Y)$ denotes the pixel arrangement with distributions; $X \in \{1,2,...,A\}$ and $Y \in \{1,2,...,B\}$, If we consider that the image has L number of thresholds, then the total thresholds of the image will be; $T = \{0,1,2,...,L-1\}$ and the following expression will relate the image dimension with the threshold distribution;

$$f(X,Y) \in T \forall (X,Y) \in Grayscale\ image \tag{1}$$

The regulated grayscale image is depicted as $R = \{r_0, r_1,...,r_{L-1}\}$
In multi-threshold operation, the above equation can be expressed as;

$$R(T) = r_0(t_1) + r_1(t_2),...,+r_{L-1}(t_{k-1}) \tag{2}$$

$$T_{max} = \max_T\{R(T)\} \tag{3}$$

Where, Eqn (3) depicts the optimal value of the SE, which is to be maximized based on the chosen threshold value during the multi-thresholding operation. In this work, $T_{max}$ for the chosen test image with a dimension $AxB$ is achieved using the FA. This work implemented a Brownian walk assisted FA as discussed in [31].
The expression of the FA implemented in this work is depicted below;

$$F_i^{t+1} = F_i^t + \beta_0 e^{-\gamma D_{ij}^2}(F_j^t - F_i^t) + \alpha_1.sign-1/2) \oplus A|s|^{\alpha/2} \tag{4}$$

where $F_i^{t+1}$ =revised location of i$^{th}$ firefly, $F_i^t$ =initial location of i$^{th}$ firefly, $\beta_0 e^{-\gamma D_{ij}^2}(F_j^t - F_i^t)$ = attraction among fireflies, $A$ =random variable, $\beta$ and $\alpha$ are the spatial and temporal exponents, respectively.

In this work, the FA is employed to identify $T_{max}$; with the help of a heuristic search operation. The FA parameters are assigned as discussed by Rajinikanth et al. [31] and the tuned FA is allowed to randomly vary the threshold values until the maximized SE is attained.

### 2.3 Image post-processing

The performance of the image based disease evaluation depends mainly on the outcome of the post-processing technique, which helps to extract the pneumonia infected section from the lung CTI. In this work, the Markov-Random-Field (MRF) segmentation technique discussed by the work of Palani et al. [30] and Rajinikanth et al. [31] is adopted. This procedure consists of the morphology based enhancement and clustering assisted segmentation of the image. This procedure works based on the MRF-Expectation Maximization (MRF-EM) idea and the information on this procedure is clearly discussed in the above said works.

Let, a grayscale image has the following representation;

$$Image = I(X,Y) \big| 0 \leq I \leq L-1 \tag{5}$$

where I represents the intensity of the image with pixel distribution $(X,Y)$. During the segmentation operation, MRF will estimate the pattern of every pixel by grouping it into a set of random variables; like $Z = \{z_1, z_2, ..., z_N\} \big| z_i \in I$.

In the implemented work, the essential labels for the MRF-EM segmentation is assigned as three, which segregated the image into background, normal section and COVID-19 infection.

In this work, the assigned labels are arranged as follows;

$$k_1 = I(X,Y) = z_1 \big| 0 \leq I \leq t_1$$
$$k_2 = I(X,Y) = z_2 \big| t_1 \leq I \leq t_2 \tag{6}$$
$$k_3 = I(X,Y) = z_3 \big| t_2 \leq I \leq t_{L-1}$$

In MRF-EM technique, the segmentation process initially enhances the image by minimizing the energy function and then implements a segmentation technique; which separates the enhanced image into three sections. Other essential information on MRF-EM segmentation can be found in [30,31].

### 2.4 Validation of proposed system

During the medical data assessment, the merit of the computerized disease inspection systems are usually authenticated based on the performance measures attained during the testing and validation using a benchmark datasets. The regular measures considered include; Jaccard, Dice, accuracy, precision, sensitivity, specificity, and Negative Predictive Value (NPV). If attained values are better (closer to unity or nearer to 100%), then the implemented disease examination scheme is confirmed as a better procedure. Further, the clinical significance of the considered scheme is to be validated using the medical data attained from real patients. In the proposed work also, similar procedure is followed to authenticate the performance of the proposed COVID-19 lesion detection system. Initially, the developed scheme is tested using the benchmark dataset existing with the GTI and later, its performance is tested by considering the clinical grade CTI collected from COVID-19 infected patients.

The mathematical expressions are depicted below [31-33];

$$Jaccard = GI \cap SI / GI \cup SI \tag{7}$$

$$Dice = F1Score = 2(GI \cap SI) / |I| \cup SI \tag{8}$$

$$Accuracy = \frac{T_{+ve} + T_{-ve}}{T_{+ve} + T_{-ve} + F_{+ve} + F_{-ve}} \tag{9}$$

$$Pr\,ecision = \frac{T_{+ve}}{T_{+ve} + F_{+ve}} \tag{10}$$

$$Sensitivity = \frac{T_{+ve}}{T_{+ve} + F_{-ve}} \tag{11}$$

$$Specificity = \frac{T_{-ve}}{T_{-ve} + F_{+ve}} \tag{12}$$

$$Negative\,Predicted\,Value(\,NPV\,) = \frac{T_{-ve}}{T_{-ve} + F_{-ve}} \tag{13}$$

where *GI*=ground-truth-image, *SI*=segmenteCOVID-19 lesion, $T_{+ve}$, $T_{-ve}$, $F_{+ve}$ and $F_{-ve}$ denotes true-positive, true-negative, false-positive and false-negative correspondingly.

## 3. Results

This section of the research depicts the investigational outcome attained using the proposed scheme. This scheme is implemented using following workstation; Intel i5 2.GHz processor with 8GB RAM and 2DB VRAM equipped with the MATLAB software. Experimental results of this study confirm that this scheme requires a mean time of 41 sec to process a CTI image with a dimension of 512x512x1 pixels. The advantage of this scheme is that it is a fully automated practice and requires the operator's assistance only for; (i) Preliminary tuning of the FA parameters and (ii) Fixing the number of iterations for the MRF-EM segmentation process.

The proposed scheme is initially tested using the benchmark images available in [28]. The chosen COVID-19 CTI and the related results attained with the proposed scheme is depicted in Figure 4. Figure 4(a) and (b) presents the chosen test image and artifact removed image respectively. Figure 4(c) shows the threshold image with FA+SE. Figure 4 (d) and (f) presents the initial and final enhancement with the MRF-EM technique, and the convergence of the Expectation Maximization (EM) search for a chosen iteration (3 trials) is depicted in Figure 4(e). Figure 4(g) and (h) depicts the segmented background and the infection section and the smoothened binary form of the infection is presented in Figure 4(i). Finally, Figure 4(j) depicts the GTI of the test image shown in Figure 4(a). Finally, a relative assessment between Figure 4(i) and (j) is executed and the essential performance measures discussed in sub-section 3.4 are computed and by using these values, a confusion matrix is formed as illustrated in Figure 5. For the chosen CTI, the proposed method helped to achieve better value of Jaccard (87.43%) and Dice (93.29%). Further, the segmentation accuracy achieved with the proposed method is 98.48%.

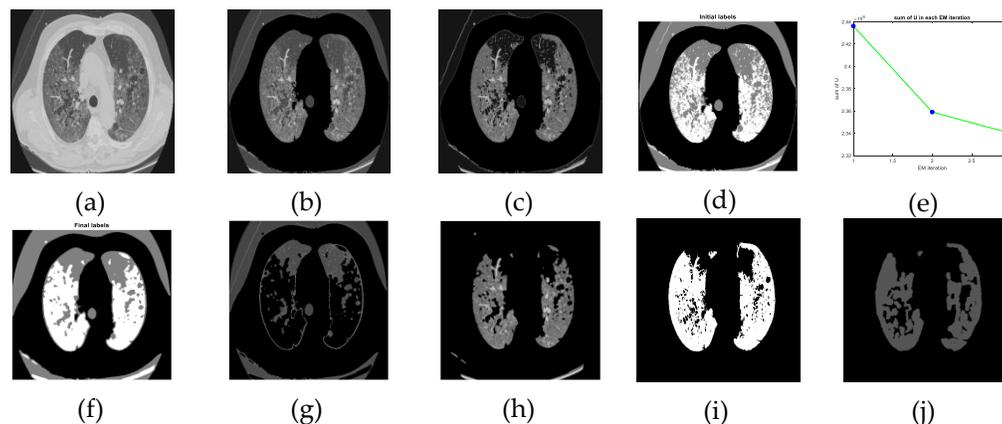

**Figure 4.** Results attained with the benchmark COVID-19 pneumonia CTI. (a) Trial image, (b) Image without artifact, (c) Threshold image, (d) Initial label, (e) Convergence of EM search, (f) Final label, (g) Extracted background, (h) Extracted Covid-19 infection, (i) Binary form of infection, (j) GTI

| True Positive= 24988 | False Negative=2527 | Sensitivity=90.82% |
|---|---|---|
| False Positive=1066 | True Negative= 208076 | Specificity=99.49% |
| Precision=95.91% | Negative Predicted Value=98.80% | Accuracy=98.48% |

**Figure 5.** Confusion matrix obtained with the chosen test image

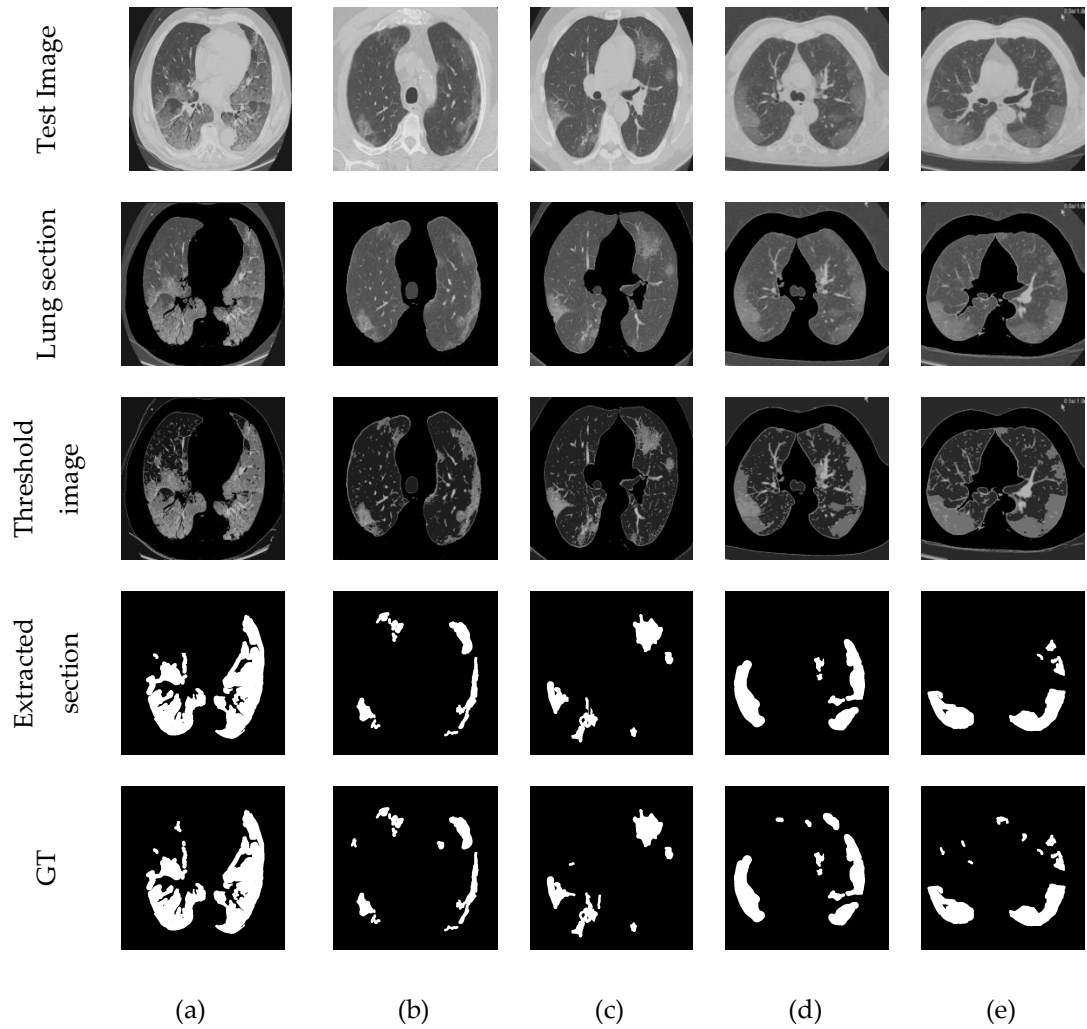

(a)          (b)          (c)          (d)          (e)

**Figure 6.** Experimental outcome attained with sample trial images

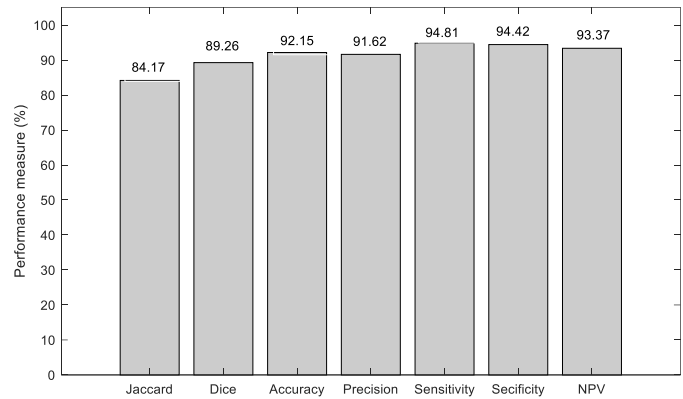

**Figure 7.** Mean performance value attained for benchmark COVID-19 CTI

Figure 6 (a)-(e) presents other test images considered in the study and the related results. This result also confirms that, proposed method is efficient in extracting the COVID19 with better accuracy. Similar procedure is then executed using other test images of the benchmark dataset considered in the proposed work, and the mean value of the performance measures obtained is considered for the validation process. The values depicted in Figure 7 confirm that, proposed procedure offered better performance measure values.

The clinical importance of the proposed system is verified using the clinical grade CTI of the Radiopaedia database. In this work, the COVID-19 pneumonia images of three patients are considered for the assessment and from every patient, 15 numbers of 2D slices are extracted for the assessment. This work considered 45 grayscale images for the assessment and the attained result of this work confirmed that, proposed practice helps to segment the COVID-19 lesions with better accuracy.

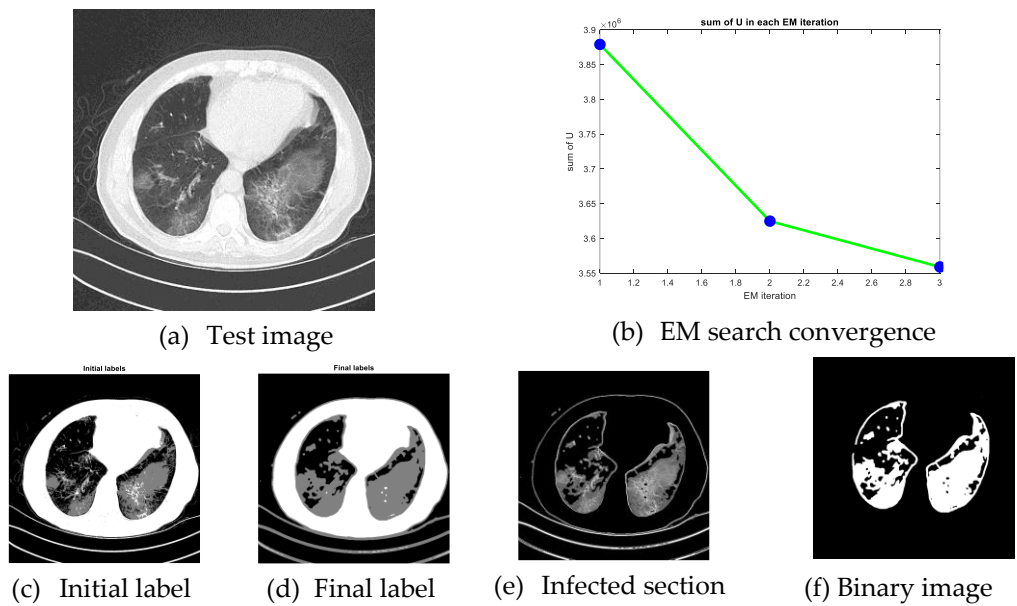

(a) Test image  (b) EM search convergence

(c) Initial label  (d) Final label  (e) Infected section  (f) Binary image

**Figure 8.** Results obtained for the Rediopaedia image.

The result attained for a sample test image is depicted in Figure 8. Figure 8(a) presents the chosen trial image; Figure 8(b) depicts the convergence of the EM search, Figure 8(c) and (d) depicts the initial and the final labels. Figure 8(e) and (f) presents the extracted infected section and its smoothened binary image respectively. Similar procedure is applied on other lung CTI and the attained result is confirmed with the physical verification. The experimental outcome confirms that, the proposed image examination offered superior results on the considered images of the Radiopaedia case studies of Case20 [34], Case22 [35], and Case25 [36].

## 4. Discussions

Starting from the date of its origin in Wuhan-China, COVID-19 has led to a major health crisis throughout the world and caused a considerable numbers of the infection as well as the morbidity rate [4]. One of the unusual features of the SARS-CoV-2 infection is that the virus infects both the upper and the lower respiratory tract. Most of the COVID-19 infected patients exhibit symptoms such as fever, dry cough and dyspnoea. Bilateral ground-glass opacities have been commonly reported in the CT scans of the COVID-19 infected patients. This specific infection has exhibited an unusually high level of pathogenicity leading to excessive lung damage [37].

With the number of infections rising every day, a need for the rapid testing of the respiratory specimens and a computer based automated analysis of the CT scans need to be performed in routine. Due to the high burden on the health care system, systems for automated analysis and screening need to be rapidly employed in health care facilities. In this regard, a need seems to have arisen for the identification and testing of effective algorithms for the automated segmentation and processing of CT scan images. Since such procedures have an implication towards the reduction in the burden of healthcare workers, researchers need to rapidly test and ascertain various algorithms which have high efficiency towards extractions of lesions formed due to COVID-19.

At present, several countries have implemented a "lockdown" intervention in the form of restrictions on travel and gatherings. This scenario has already posed a huge burden on the economy in addition to the already existing burden on the healthcare system due to the high case loads. In this regard, a mass screening facility for the screening of CT images as proposed in this research work will have a huge impact in terms of the reduction of this burden.

## 5. Conclusion

The objective of this research is to propose an automated image examination scheme to detect and extract the COVID-19 lesions from the lung CT scan images. This work implemented a sequence of procedures to extract the COVID-19 lesion with greater accuracy. This work employed the image processing methods, such as artifact removal, FA and SE based three-level thresholding to enhance the image, MRF-EM segmentation to extract the COVID-19 lesion and a relative assessment of the extracted binary image with the GTI to compute the performance measures. This work initially considered the 2D CTI of benchmark COVID-19 dataset as well as the clinical-grade 2D slices for the experimental investigation. The initial result attained with the benchmark dataset offered a mean segmentation accuracy of >92% on 50 images of dimension 512x512x1 pixels. The results attained with the clinical grade CTI also offered better segmentation of the COVID-19 lesion. The experimental results attained from this study confirmed the superiority of the proposed scheme and in future, this scheme can be used to examine the clinically obtained lung CTI of COVID-19 patients.

**Author Contributions:** Conceptualization, Data curation, methodology, formal analysis, writing—original draft preparation, validation, V.R.; Conceptualization, writing—original draft preparation, supervision, validation, writing—review and editing, project administration ,S.K; software, validation,

K.P.T; writing—original draft preparation, validation, K.K. All authors have read and agreed to the published version of the manuscript.

**Funding:** This research received no external funding

**Acknowledgments:** The authors of this paper would like to thank Medicalsegmentation.com and Radiopaedia.org for sharing the clinical grade COVID-19 images.

**Conflicts of Interest:** The authors declare no conflict of interest.